\documentclass{article}
\usepackage[utf8]{inputenc}

\usepackage{amsmath,amsfonts}
\usepackage{amssymb}
\usepackage{geometry}
\usepackage{float}
\usepackage{comment}
\usepackage[normalem]{ulem}
\usepackage{multirow}
\useunder{\uline}{\ul}{}
\usepackage[toc,page]{appendix}
\usepackage{cite}
\usepackage[pdftex]{graphicx}
\usepackage{algorithmic}
\usepackage{array}
\usepackage{fixltx2e}
\usepackage{subfigure}
\usepackage{lscape}
\usepackage[autostyle]{csquotes}
\usepackage{hyperref}
\usepackage{atbegshi}
\usepackage{fancyhdr}

\usepackage[numbers]{natbib}

\title{Fusing Low-Latency Data Feeds with Death Data to Accurately Nowcast COVID-19 Related Deaths\footnote{\bf{This research was presented at the Joint Statistical Meetings (JSM) 2021: Statistics, Data, and the Stories They Tell and subsequently published as part of the proceedings.}}
}

\author{Conor Rosato\thanks{Department of Electrical Engineering and Electronics, University of Liverpool, United Kingdom} \and  Robert E. Moore$^\dagger$ \and  Matthew Carter$^\dagger$ \and John Heap\thanks{Computational Biology Facility, University of Liverpool, United Kingdom}\and Jose Storopoli\thanks{Department of Computer Science, Universidade Nove de Julho - UNINOVE, Sao Paulo - Brazil} \and Simon Maskell$^\dagger$}

\begin{document}

\maketitle

\begin{abstract}

The emergence of the novel coronavirus (COVID-19) has generated a need to quickly and accurately assemble up-to-date information related to its spread. While it is possible to use deaths to provide a reliable information feed, the latency of data derived from deaths is significant. Confirmed cases derived from positive test results potentially provide a lower latency data feed. However, the sampling of those tested varies with time and the reason for testing is often not recorded. Hospital admissions typically occur around 1-2 weeks after infection and can be considered out of date in relation to the time of initial infection. The extent to which these issues are problematic is likely to vary over time and between countries. 

We use a machine learning algorithm for natural language processing, trained in multiple languages, to identify symptomatic individuals derived from social media and, in particular Twitter, in real-time. We then use an extended SEIRD epidemiological model to fuse combinations of low-latency feeds, including the symptomatic counts from Twitter, with death data to estimate parameters of the model and nowcast the number of people in each compartment. The model is implemented in the probabilistic programming language Stan and uses a bespoke numerical integrator. We present results showing that using specific low-latency data feeds along with death data provides more consistent and accurate forecasts of COVID-19 related deaths than using death data alone.

\end{abstract}

\section{Introduction}\label{intro}

The novel coronavirus (COVID-19), has at the time of writing, resulted in over 4.55 million deaths and 219 million confirmed cases worldwide as of 6th October 2021. By January 2020, new cases had been seen throughout Asia, and by the time the World Health Organisation (WHO) declared a global pandemic in March 2020, COVID-19 had spread to over 100 countries. Therefore, it was imperative to establish reliable data feeds relating to the pandemic so that researchers and analysts could model the ongoing spread of the disease and inform decision-making by government and public health officials. These data sets and models must be open-source to facilitate collaboration between researchers and allow for published results to be replicated and scrutinised. A popular interactive dashboard that collates total daily counts of confirmed cases and deaths for countries, and in some cases, regions within countries exists here \cite{DONG2020533}. These variables are traditionally used to calculate metrics such as the reproduction number ($R_{t}$), which is vital in understanding both the number of people on average an infected person infects and the infection growth rate or daily rate of new infections. The quality of the metrics calculated is heavily dependent on the model and ingested data. 

In the United Kingdom (UK), there has been a joint effort to produce estimates of the ($R_{t}$) number, with some notable examples seen here \cite{rt}. Laboratory-confirmed COVID-19 diagnoses are used in \cite{cambridge_model}, UK's National Health Service (NHS) Pathways data is used in \cite{exeter_model} and hospital admissions data is used in \cite{warick}. The statistical model developed by Moore, Rosato and Maskell \cite{moore_assessing_2022} contributed to these estimates by incorporating death, hospital admission and NHS 111 call data.

Terms such as \enquote{Infodemiology}, and \enquote{Infoveillance} described in \cite{infovelliance} refer to the ability to process and analyse data that is created and stored digitally in real-time pertinent to disease outbreaks. The availability of these datasets, particularly at the beginning of an outbreak when very little is known, could provide a noisy but accurate representation of disease dynamics. A popular method includes extracting data from social media and, in particular, Twitter. Before the pandemic, tweets relating to influenza-like-illness symptoms were seen to substantially improve the models predicting capacity in \cite{tweets_6} and boost nowcasting accuracy by 13\% in \cite{Sentinel_1}. In relation to the COVID-19 pandemic, there have been many research papers published that use social media to gain valuable information relating to the pandemic from what people tweet in real-time. Public sentiment relating to prevention strategies was analysed in \cite{tweets} while \cite{tweets_1} showed that emotion changed from fear to anger during the first stages of the pandemic. Misinformation and conspiracy theories have been shown to have propagated rapidly through the Twittersphere during the pandemic \cite{tweets_2}. Studies have used machine learning algorithms to automatically detect tweets containing self-reported symptoms mentioned by users \cite{tweets_3} with \cite{tweets_4} finding that the symptoms reported by Twitter users were similar to those used in a clinical setting. To the best of our knowledge, researchers have yet to use these symptomatic tweets to calibrate epidemiological models.

The contribution of this paper is twofold: firstly, we outline how to identify symptomatic tweets that correspond to COVID-19 related symptoms in multiple languages. The geolocation information associated with each tweet, when available, is extracted, and counts per country or region are aggregated to produce estimates for the previous 24 hours. Secondly, we present a comprehensive study of how these symptomatic tweets differ from other open-source datasets when used to calibrate the extended SEIR model described in section \ref{sec:model} for different geographic locations. When incorporating the surveillance data, outlined in section \ref{sec:surveillance_data}, the Mean Absolute Error (MAE) and Normalised Estimation Error Squared (NEES) values are calculated when making 7-day death forecasts.

An outline of the paper is as follows. In Section \ref{sec:surveillance_data} we describe the open-source data feeds included in the comparative study and provide the methodology for extracting the symptomatic tweets in real-time. A description of the model is presented in Section \ref{sec:model}, with an outline of the computational experiments and results in Section \ref{sec:computational_experiments}. Concluding remarks and directions for future work are described in Section \ref{sec:conclusions_future_work}.

\section{Data\label{sec:surveillance_data}}

The surveillance data used for each geographical location are summarised in Table \ref{table:data_feeds}. Death and positive case data for the US States and the rest of the world were downloaded from the dashboard operated by the Johns Hopkins University Center for Systems Science and Engineering (JHU CSSE) \cite{DONG2020533}. It should be noted that testing methods and criteria for classifying deaths as COVID-19-related differ between geographic locations.

For NHS region-specific data, the number of deaths includes individuals with COVID-19 as the cause of death on their death certificate or those who died within 60 days of a positive test result. Patients admitted to hospital with COVID-19 symptoms and individuals that input symptoms to the ZOE COVID Symptom Study app comprise the hospital admissions and Zoe app datasets, respectively. Individuals that reported symptoms via the NHS Pathways triage and online Dashboard comprise the 111 calls and 111 online assessments datasets, respectively. Note that the Zoe app, 111 calls and 111 online assessments may include individuals who have COVID-19 symptoms but have not tested positive and individuals who perceive they have symptoms and do not have COVID-19.

All code and datasets can be found on the CoDatMo GitHub repository \cite{codatmo}. The authors set this up to facilitate the sharing of code, data and ideas when modelling COVID-19. 


\begin{table}
\scriptsize
\caption{Data Sources}
\begin{center}
\begin{tabular}{cccc}
\hline
\hline
\\[-5pt]
\multicolumn{1}{c}{\textbf{Geographic Location}} &
\multicolumn{1}{c}{\textbf{Data Feed}} &
\multicolumn{1}{c}{\textbf{Start Date}} &
\multicolumn{1}{c}{\textbf{Reference}} \\
&         &  \\
\hline
U.S States and the rest of the world &      Deaths  & 24th March 2020 & \cite{DONG2020533}  \\
                         &      Tests   & 1st March 2020 & \cite{DONG2020533}   \\
                         &      Twitter &  13th April 2020 &Section \ref{sec:twitter}   \\

\hline
U.K NHS Regions          &      Deaths  & 24th March 2020 & \cite{deaths}  \\
                         &      Hospital admissions & 19th March 2020 & \cite{hospital}   \\
                         &      Twitter    & 9th April 2020 & Section \ref{sec:twitter} \\ 
                         &      Zoe app    & 12th May 2020 & \cite{zoe_app}   \\
                         &      111 calls  & 18th March 2020 & \cite{calls_online}  \\
                         &      111 online & 18th March 2020 & \cite{calls_online}    \\
\hline
\end{tabular}
\end{center}
\label{table:data_feeds}
\end{table}

\subsection{Twitter\label{sec:twitter}}

We created an interactive website\footnote{https://pgb.liv.ac.uk/$\sim$johnheap/} that maps symptomatic tweets to geographical locations with daily counts representing the total amount of symptomatic tweets from the previous 24 hours. Information on how to download the data as a JSON can be found on the website. 

\subsubsection{Pre-processing}

The Twitter streaming API is filtered using keywords that align with COVID-19 symptoms from the MedDRA database \cite{medra} in English, German, Italian, Portuguese and Spanish, including terms for fever, cough and anosmia. We note that our analysis indicated that explicit COVID-19 terms (e.g. `coronavirus') rarely related to individuals with symptoms. Such terms were therefore excluded. Official retweets or tweets beginning with \#RT were removed to avoid duplication of tweets within the dataset.

\subsubsection{Symptom Classifier Breakdown}

A multi-class support vector machine (SVM) \cite{scikit-learn} was trained with a set of annotated tweets that were vectorised using a skip-gram model. The annotated tweets were labelled according to the following classes:

\begin{enumerate}
\item Unrelated tweet,
\item User currently has symptoms,
\item User had symptoms in the past,
\item Someone else currently has symptoms,
\item Someone else had symptoms in the past.
\end{enumerate}

\noindent The sum of tweets in classes 2-5, which is the total number of tweets that mention symptoms, was calculated for each 24-hour period. Geo-tagged tweets were mapped to their location, e.g. the corresponding city, town or village, via a series of tests using shapefiles of different countries. Previous studies demonstrate that approximately 1.65\% of tweets are geo-tagged \cite{tweets_5}, where the exact position of the tweeter when the tweet was posted is recorded using longitude and latitude measurements. For tweets that are not geo-tagged, we look at the author's profile to ascertain whether they provide an appropriate location. We deemed the server offline if there were any 15 minute periods during the previous 24 hours that did not have any recorded tweets. After checking all 96 15 minute periods, the count in each geographical area was multiplied by a correction factor: reported tweet count = total tweet count * 96/(96 - downtime periods). 

For each language, we labelled the corpus of tweets with native speakers, with the associated class label and randomly up- and down-sampled under- and over-represented classes such that the classifier was trained with a balanced dataset. A subset of data was used to train the classifier before testing it on the remainder. The total number of labelled tweets used for training and testing of the classifiers and the resulting performance metrics can be seen in Table \ref{table:performance}.

\begin{table}
\scriptsize
\caption{Performance Measures}
\begin{center}
\begin{tabular}{c|cc|cccc}
\hline
\hline
\\[-9pt]
\multicolumn{1}{c}{\textbf{Language}} &
\multicolumn{2}{|c|}{\textbf{Number of Data Used}} & &
\multicolumn{2}{c}{\textbf{Performance Measures}} & 
\multicolumn{1}{c}{}\\
\multicolumn{1}{c}{} &   
\multicolumn{1}{|c}{Training} &
\multicolumn{1}{c|}{Testing} &
\multicolumn{1}{c}{F1} &
\multicolumn{1}{c}{Accuracy}&
\multicolumn{1}{c}{Precision} &
\multicolumn{1}{c}{Recall}\\
\hline
English&  1105& 195 & 0.85 & 0.85 & 0.85 & 0.85  \\
German&  412 & 260&     0.89 & 0.89 & 0.90 & 0.89  \\
Italian& 254 & 260 &     0.97 & 0.96 & 0.97 & 0.96  \\
Portuguese& 3507 & 619 & 0.77 & 0.77 & 0.78 & 0.80 \\
Spanish&  1530 & 270&    0.82 & 0.85 & 0.82 & 0.85  \\
\hline
\end{tabular}
\end{center}
\label{table:performance}
\end{table}

\section{Model\label{sec:model}}

We repurpose the statistical model developed by Moore, Rosato and Maskell~\cite{moore_assessing_2022} by tweaking the observation model to be compatible with each group of surveillance data types that we use to calibrate the model in the computational experiments. We calibrate the model with a minimum of death data in all experiments, and the associated component of the observation model is unchanged. We extend the observation model to assimilate the other types of surveillance data that feature in Table~\ref{table:data_feeds}, including Twitter and Zoe app data, by adding an extra component for each additional data type. These extra components of the observation model, the number of which can change between experiments, mirror the structure for symptom report data in the original model. More explicitly, we assume for these extra components that a generic count on day $t$, $x_\mathrm{obs}\left(t\right)$, has a negative binomial distribution,

\begin{equation}
    x_{\mathrm{obs}}\left(t\right) \sim \mathrm{NegativeBinomial}\left(x\left(t\right), \phi_{x}\right),
\end{equation}

\noindent parameterised by a mean $x\left(t\right)$ and overdispersion parameter $\phi_x$.

\section{Computational Experiments\label{sec:computational_experiments}}

The time series we consider begins on 17th February 2020, with the start dates of the data feeds outlined in Table \ref{table:data_feeds}. We consider the end of time in our analysis for the US States and the rest of the world and NHS regions to be the 1st February 2021 and 7th January 2021, respectively. In all cases, we consider a forecast to include seven days. For US states and the rest of the world, we include three predictions in the analysis; 9th July 2020 to 16th July 2020, 17th October 2020 to 24th October 2020 and 25th January 2021 - 1st February 2021. The analysis for UK NHS regions includes six predictions; 11th November 2020 to 18th November 2020, 21st November 2020 - 28th November 2020, 1st December - 8th December, 11th December - 18th December, 21st December - 28th December and 31st December 2020 - 7th January 2021.

Similar to the experiments in \cite{moore_assessing_2022}, the analysis in this paper was run on the University of Liverpool's High-Performance Computer (HPC). Each node has two Intel(R) Xeon(R) Gold 6138 CPU @ 2.00GHz processors, a total of 40 cores and 384 GB of memory. In the following experiments, six independent Markov Chains draw 2000 samples each, with the first 1000 discarded as burn-in. Run-time differs depending on the country and at what point in the time series the prediction is made, but it typically takes 4.5 hours per Markov Chain to complete.

We initially calibrate the model solely with death data and produce posterior predictive distributions of deaths for the following geographic locations independently: 
\begin{description}
  \item[$\cdot$ US:] 50 States,
  \item[$\cdot$ Rest of World:] 2 European and 16 Latin American countries,
  \item[$\cdot$ UK:] 7 NHS regions.
\end{description}

\noindent We consider the final 7-daily deaths in this forecast to be the baseline to compare forecasts of deaths when incorporating low-latency data feeds. We use two metrics in our analysis to determine the accuracy of the resulting forecasts. Firstly, we calculate the MAE, which shows the average error over a set of predictions:

\begin{eqnarray}
\mathrm{MAE}=\frac{1}{N} \sum_{i=1}^N (x^i-y^i) \label{eqn:MAE},
\end{eqnarray}

\noindent where $N$ is the number of predictions, and $x_i$ and $y_i$ are the predicted and true number of deaths on day $i$, respectively.

Secondly, we consider the uncertainties associated with the forecasts by assessing the NEES, which is a popular method in the field of signal processing and tracking \cite{NEES} and recently applied to epidemiological forecasts in \cite{moore_assessing_2022}, to determine if the estimated variance of forecasts from an algorithm differs from the true variance. If the variance is larger than the true variance, then the algorithm is over-cautious, and if the estimated variance is smaller than the true variance, it is over-confident. The NEES is defined by:

\begin{eqnarray}
\mathrm{NEES} = \frac{1}{N} \sum_{i=1}^N  (x^i-y^i)^T {C^i}^{-1} (x^i-y^i),  \label{eqn:NEES}
\end{eqnarray}

\noindent where ${C^i}^{-1}$ is the estimated variance at day $i$, as approximated using the variance of the samples for that day. If $x^i$ is $D$ dimensional, then $C^i$ should be a $D\times D$ matrix, and the NEES should be equal to $D$ if the algorithm is consistent. Therefore, in assessing death forecasts, an ideal NEES value is $D\approx 1$. 

\subsection{Results}

The NEES values and MAE percentage differences between the baseline, of ingesting solely deaths, and the incorporation of low-latency data feeds for US States and the rest of the world and UK's NHS regions can be seen in Tables \ref{table:USstates} and \ref{table:UKNHS}, respectively. The results in these tables are averaged over the prediction periods described in section \ref{sec:computational_experiments}.

\begin{figure}[H]
\centering 
\includegraphics[scale=0.26, bb=500 0 600 600]{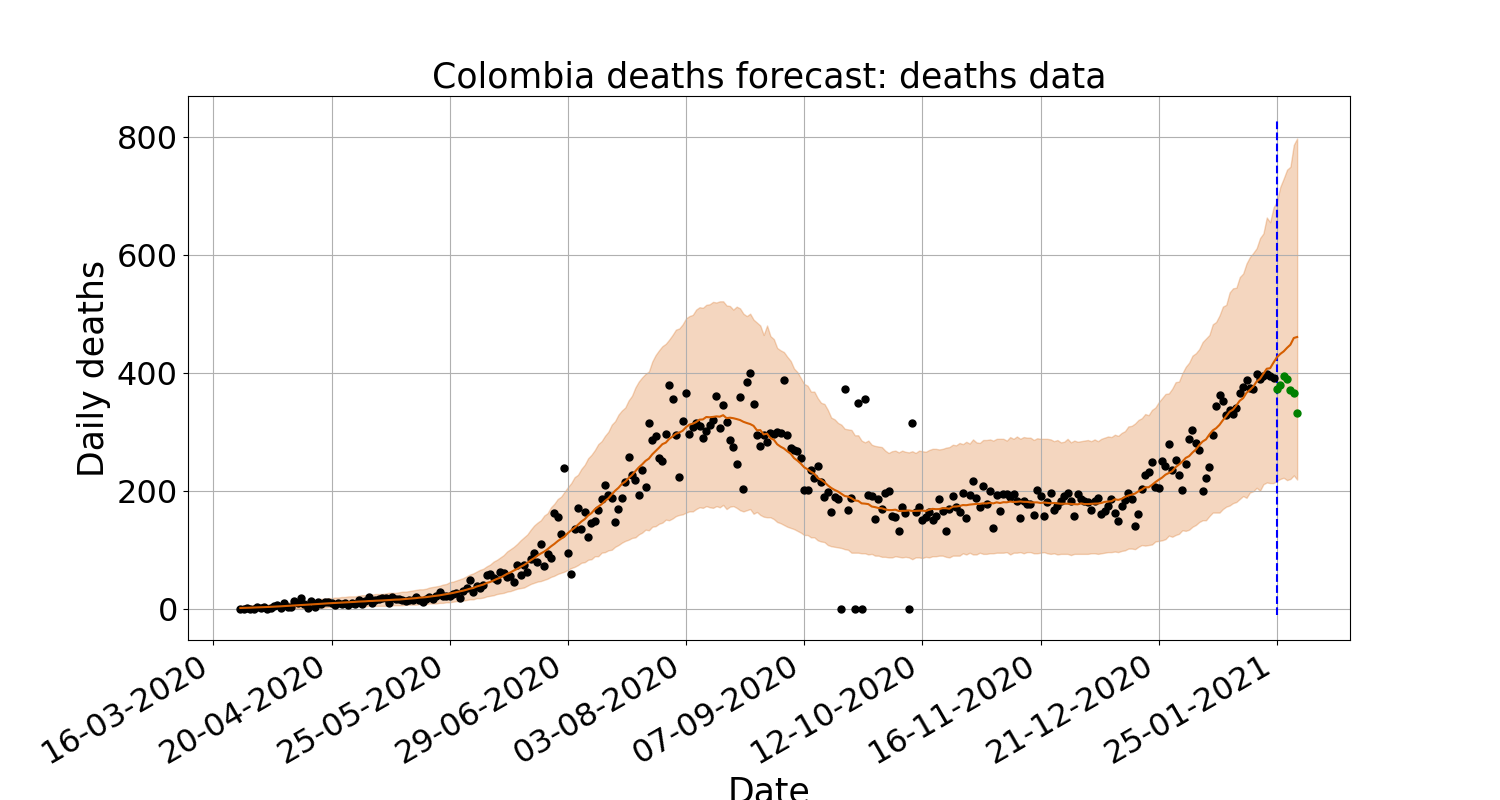} \\
\includegraphics[scale=0.26, bb=500 0 600 600]{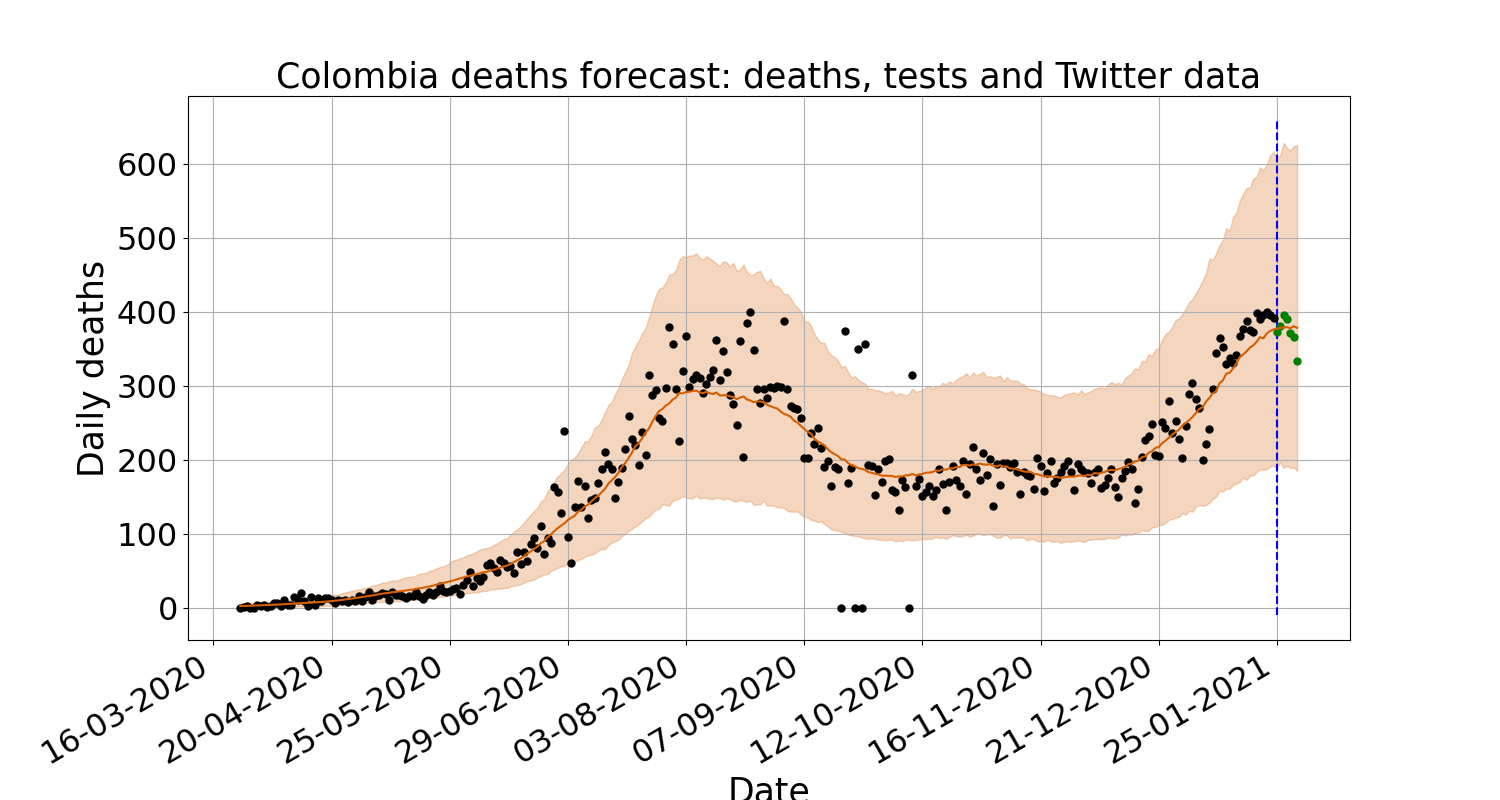}%
\caption{Deaths forecast for Colombia using solely death data (top) and deaths, tests and Twitter data (bottom). The orange ribbon is 1 standard deviation from the mean, the red line is the mean sample and the start of the predictions is the vertical dashed blue line. The black and green dots are the true deaths.}
\label{fig:mae}
\end{figure}

When forecasting deaths using the data available from \cite{DONG2020533}, we have shown that calibrating the model with tests and tweets gives comparable increases in performance for US States, however for the rest of the world, tweets give a -17\% improvement compared to just -6\% for tests. For US States and the rest of the world, there is an improvement of -5\% and -24\%, respectively when tests and tweets are used to calibrate the model. An example of this improvement can be seen in Figure \ref{fig:mae} for the prediction period 25th January 2021 - 1st February 2021. When comparing the mean sample, outlined in red, incorporating tests and tweets follows the deaths trend with more accuracy.

\begin{figure}[H]
\centering 
\includegraphics[scale=0.24, bb=500 0 600 600]{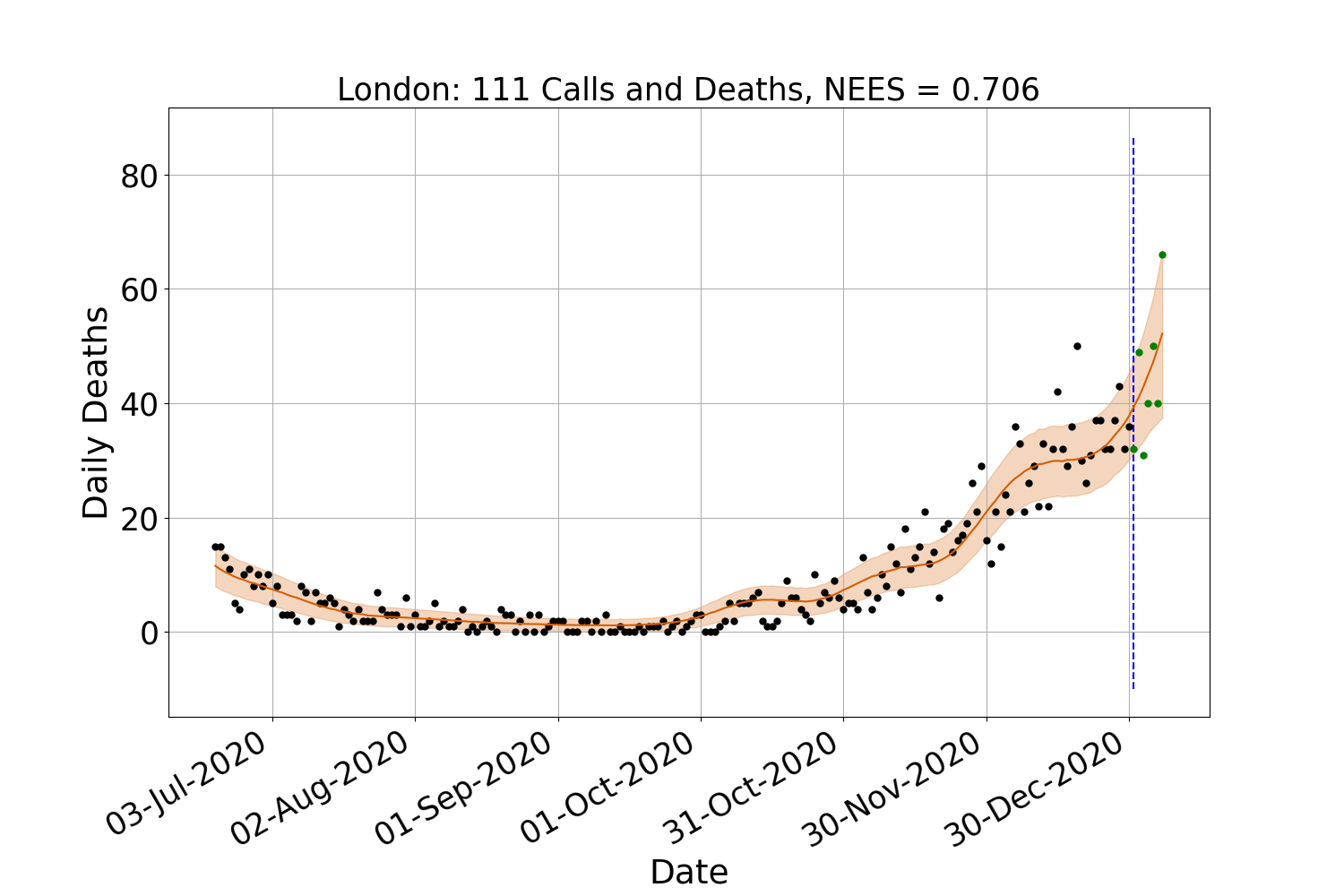} \\[1cm]%
\includegraphics[scale=0.24, bb=500 0 600 600]{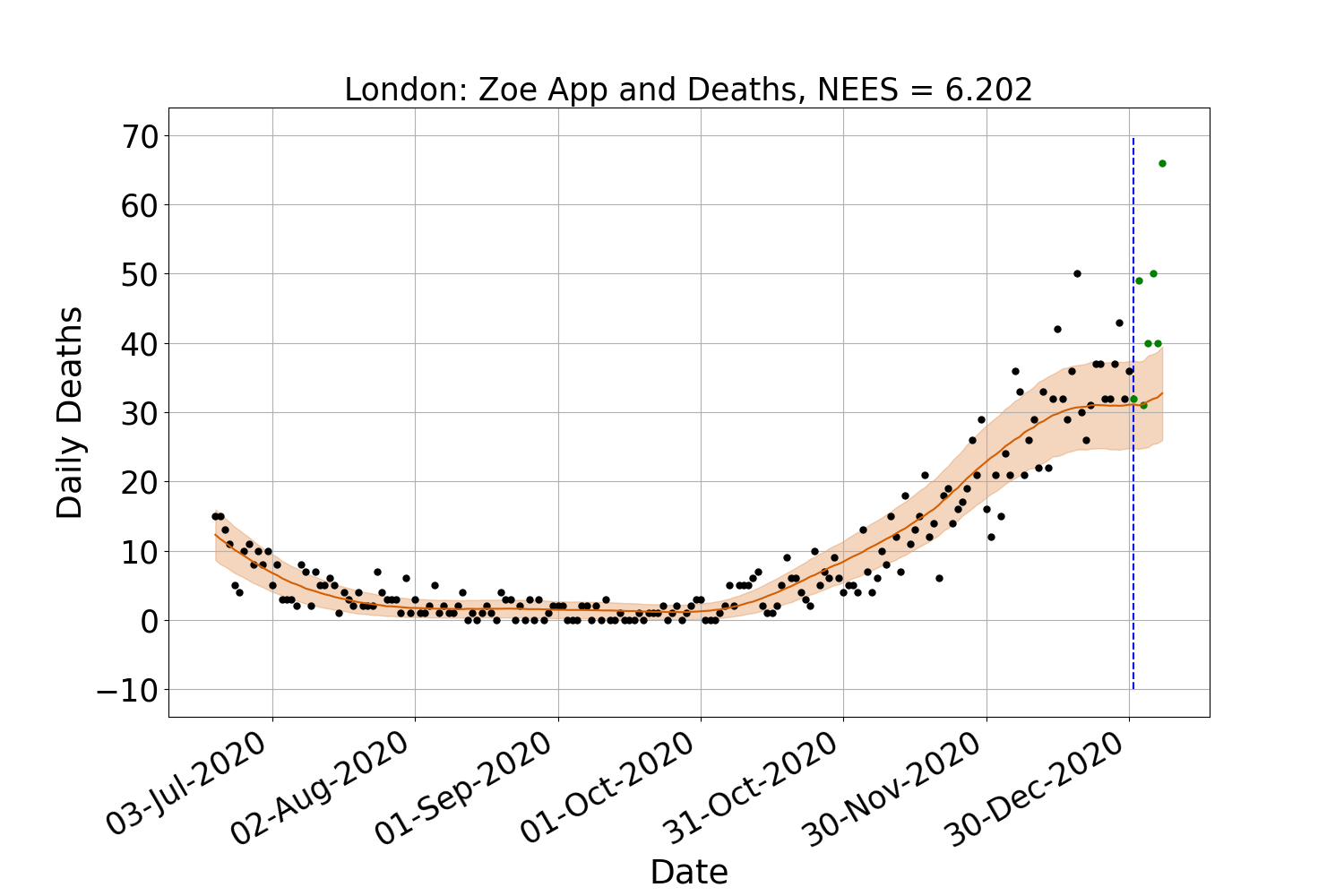}%
\caption{Deaths forecast for London using death and 111 calls data (top) and death and Zoe App data (bottom). The orange ribbon is 1 standard deviation from the mean, the red line is the mean sample and the start of the predictions is the vertical dashed blue line. The black and green dots are the true deaths.}
\label{fig:London}
\end{figure}

When comparing the NEES values for NHS regions in Table \ref{table:UKNHS} against the baseline, it is evident that including a data feed improves the consistency of forecasts. The exception is the Zoe App data, which on average, produces estimates that are over-confident. This can be seen in Figure \ref{fig:London} for the prediction period 31st December 2020 - 7th January 2021. When comparing the MAE \% difference, the results are similar. Ingesting hospital admissions, 111 calls, and 111 online assessments data provides improvements of -22\%, -17\% and -22\%, respectively. However, there is a more significant difference in the MAE of 124\% when including the Zoe App data than the 2\% from including the Tweet data. We perceive this issue arises because, in the context of these feeds, the symptoms are self-diagnosed. Consequently, the counts may well include relatively large numbers of people who do not have COVID-19. We do not currently consider such `false alarms' in the model described in section \ref{sec:model} but hope to extend it to handle these in the future.

\section{Conclusions and Future Work\label{sec:conclusions_future_work}}

We have shown that calibrating the epidemiological model outlined in section \ref{sec:model} with certain low-latency data feeds provides more accurate and consistent nowcasts of daily deaths when compared with using death data alone. 

Incorporating tweets for UK regions does not provide the same level of improvement as for other geographies. This reduced improvement could be down to many factors, including the total daily counts for NHS regions being less plentiful than for the US States or the rest of the world. We used the free Twitter streaming API for this research, which limits the number of tweets available to download to 1\%. However, it is possible to pay for a Premium API that allows the user to download a higher percentage of tweets. A second way to potentially increase the hit rate of geo-located tweets is to use natural language processing techniques, such as those outlined in the review \cite{NLP}, to estimate the location of the tweet user. Another direction for future work is to train a more sophisticated classifier such as the Bidirectional Encoder Representations from Transformers (BERT) classifier \cite{BERT}.

As for the symptom report data in the original model that Moore, Rosato and Maskell introduce~\cite{moore_assessing_2022}, we have assumed that all types of low-latency surveillance data are a weighted sum of current and lagged instances of the time series of new infections. We want to consider alternative structures for the observation model that link directly to the intermediate states of the transmission model in the future.

\section*{Funding}

This work was supported by a Research Studentship jointly funded by the EPSRC and the ESRC Centre for Doctoral Training on Quantification and Management of Risk and Uncertainty in Complex Systems Environments [EP/L015927/1]; an ICASE Research Studentship jointly funded by EPSRC and AWE [EP/R512011/1]; the EPSRC Centre for Doctoral Training in Distributed Algorithms [EP/S023445/1]; and the EPSRC through the Big Hypotheses grant [EP/R018537/1].

\section*{Acknowledgements}

The authors would like to thank Serban Ovidiu and Chris Hankin from Imperial College London, Ronni Bowman, Riskaware and John Harris for their support and helpful discussions of this work. We would also like to thank the team at the Universidade Nove de Julho - UNINOVE in Sao Paulo, Brazil with the help they provided in labelling the Portuguese tweets. We would also like to thank Breck Baldwin for helping to progress CoDatMo.


\bibliographystyle{unsrt}
\bibliography{bibliography.bib}

\begin{table}
\label{Table:Results}
\tiny
\caption{US States and the rest of the world: nowcasting mean absolute error per geographic location when using only death data and when using deaths and different low-latency data feeds.}
\begin{center}
\begin{tabular}{cccccccccccc}
\hline
\hline
\\[-5pt]
\multicolumn{1}{c}{\textbf{Geographic Location}} & 
\multicolumn{1}{c}{\textbf{Deaths}} & 
\multicolumn{2}{c}{\textbf{Tests}} &
\multicolumn{2}{c}{\textbf{Twitter}} &
\multicolumn{2}{c}{\textbf{Tests and Twitter}} &
\multicolumn{1}{c}{}\\
\multicolumn{1}{c}{} &    
\multicolumn{1}{c}{NEES} &  
\multicolumn{1}{c}{MAE \% Diff}&
\multicolumn{1}{c}{NEES} &
\multicolumn{1}{c}{MAE \% Diff} &
\multicolumn{1}{c}{NEES} &
\multicolumn{1}{c}{MAE \% Diff} &
\multicolumn{1}{c}{NEES} \\
\hline
\textbf{U.S States}    &   &  &   \\
Alaska& 0.329  & -36  & 0.334 &  -29 & 0.301 & -92 & 0.302\\
Alabama&  0.684    & -29  & 1.874 & -29  & 1.723 & -2 & 1.000\\ 
Arkansas&  0.275    &  3 & 0.317  & -1 & 0.288 & -1 & 0.313\\
Arizona&0.337    &  20 & 0.334 & 18 & 0.344 & -20 & 0.244\\
California& 0.611  & 6  & 0.709 & 9 & 0.802 & 5 & 1.206\\
Colorado& 1.886   & -25  & 0.401 & -41  & 0.457 & 10 & 1.278\\
Connecticut& 13.406   & -8  & 1.922 & -2 & 0.875 & 2 & 11.459\\
Delaware& 3.020  &  -3 & 0.918 & 16 & 1.046 & 12 & 0.727\\
Florida& 0.406    & -24  & 0.179 &13 & 0.353 & -20 & 0.454\\
Georgia& 0.550  & 9  & 0.325 & 41 & 0.891 & -48 & 0.255\\
Hawaii& 11.459  & -12  & 28.114 & -4 & 24.695 & 17 & 10.149\\
Iowa& 19.176  & 5  & 7.720 & 4 &1.476 & -3 & 1.600\\
Idaho& 0.914  &  0 & 0.809 & 2 &1.791 & 7 & 0.986\\
Illinois& 0.573  & 9  & 0.350 & 13 &0.319 & -116 & 1.091\\
Indiana&0.561  & -17  & 0.652 & -40 &0.781 & 0 & 0.481\\
Kansas& 1.021 &  1 & 1.037 & -2 &1.835 & 1 & 0.488\\
Kentucky& 0.355 & -4  & 0.374 & 10 &0.548 & -15 & 0.214\\
Louisiana& 0.298  & -7  & 0.305 & -2 &0.341 & 9 & 0.234\\
Massachusetts& 0.351   & 3  & 0.342 & -3 &0.365 & 14 & 0.409\\
Maryland& 0.485  &  -3 & 0.619 & 10 &0.581 & 31 & 0.313\\
Maine& 0.488  & 1 & 0.567 & -28 &0.796 & -9 & 0.952\\
Michigan& 0.592  & -6  & 0.445 & -7 &0.453 & 4 & 0.850\\
Minnesota& 0.683 & 9  & 1.019 & 11 &1.200 & 51 & 0.747\\
Missouri& 0.810 & -7  & 1.165 & -27 &1.609 & 20 & 0.475\\
Mississippi& 0.683 & 12  & 0.721 & 2 &0.997 & -15 & 0.320\\
Montana& 5.034 & 4  & 2.244 & -1 &1.538 & -5 & 5.189\\
North Carolina& 0.908   &  -1 & 0.453 & 9 &0.877 & -19 & 0.570\\
North Dakota& 0.513  & -32  & 0.521 & -18 &0.544 & -8 & 0.661\\
Nebraska& 0.259 & 5  & 0.253 & 7 &0.570 & 5 & 0.286\\
New Hampshire&0.252 & -74  & 0.240 & -148 &0.430 & -36 & 0.288\\
New Jersey& 0.901  & -7  & 0.788 & -6 &0.926 & 10 & 3.177\\
New Mexico& 0.832  & -28  & 0.738 & -12 &0.969 & 0 & 0.489\\
Nevada& 2.129 & -24  & 0.353 & -12 &0.425 & -13 & 1.904\\
New York& 0.496 & 31  & 0.146 & 3 &0.135 & -17 & 0.418\\
Ohio& 0.263 &  63 & 0.675 & 54 &0.468 & 3 & 0.337\\
Oklahoma&0.301 & -5  & 0.369 & 0 &0.621 & 8 & 0.256\\
Oregon& 0.729 & 0  & 1.032 &-2 &1.692 & -4 & 0.793\\
Pennsylvania& 0.411 & -7  & 0.385 & 0 &0.426 & 10 & 0.402\\
Rhode Island& 0.609 &  -9 & 0.546 & -31 &0.446 & -2 & 1.699\\
South Carolina& 2.072 & -3  & 2.157 & -4 &5.601 & -39 & 0.429\\
South Dakota&1.259 & 14  & 1.080  & -2 &1.089 & 2 & 5.050\\
Tennessee&0.794 & 15  & 1.191 & 14 &1.687 & -11 & 0.600\\
Texas&0.585 & 6  & 0.784 & 1 &0.750 & -71 & 0.706\\
Utah& 0.499 & -98  & 0.716  & -127 &1.196 & 13 & 0.632\\
Virginia& 0.731 & -10  & 0.396 & 6 &0.864 & 9 & 0.676\\
Vermont& 0.142 & 59  & 0.300 & -1 &0.163 & 40 & 0.043\\
Washington&0.608  & -8  & 0.561 & 19 &1.787 & -1 & 0.782\\
Wisconsin& 0.842 & 6  & 1.028 & 25 &3.921 & 8 & 0.850\\
West Virginia& 0.650 & -6  & 0.547 & 2 &1.042 & 7 & 0.291\\
Wyoming& 1.939 & 5  & 0.951 & -15 &1.126 & 25 & 0.395\\
\hline
& & & & & & \\ 
\textbf{Average}&  \textbf{1.696}  & \textbf{-5}  & \textbf{1.409} & \textbf{-6} & \textbf{1.483} & \textbf{-5} & \textbf{1.269}\\
\hline
\textbf{Rest of the World}    &   &  &  &  \\
Argentina & 0.567  & 3  & 0.695 & -17&0.904 & -19&0.765 \\
Bolivia& 0.339  & -85 & 0.207 & -117&0.182 & -118 &0.195 \\
Brazil& 0.396  & -4  & 0.405 & 11&0.578 & 4 & 0.493\\
Chile& 0.371  & 15  & 0.439 & 14&0.506 & 10& 0.425\\
Colombia& 0.154 & 17  & 0.243 & -46&0.164 & -115 & 0.223\\
Costa Rica& 0.423 & 6  & 0.583 & 18&3.060 & 2&0.786\\
Ecuador& 0.156 & -26  & 0.195 & -99&0.234 &-69 &0.234\\
Guatemala& 0.557 & -19  & 0.670 & -31&0.815 & -31&0.713\\
Honduras& 0.405 & -8  & 0.381 & -27&0.915 &-41 &0.541\\
Mexico& 0.766  & 16  & 0.939 & 11&1.100 &11 &1.110\\
Nicaragua& 0.091 & -13  & 0.207 & -24 &1.340 &-22 &0.364\\
Panama& 0.550 &  -20 & 0.421 & -4&0.451 & -7&0.368\\
Paraguay& 0.535 & 28  & 0.877 & -7&2.615&8 &1.473\\
Peru&  0.507  & 33  &0.103 & 26 &1.630 & 16&0.515\\
Uruguay& 0.619   & 11  & 0.742 & -13&0.899&-7 &0.643\\
Venezuela& 0.610   & -14  & 0.713 & -49&0.890&-91 &0.603\\
Germany& 0.379 &  5 & 0.613 & 15&2.131&14 &1.570\\
Italy& 0.360  &  17 & 0.557 & 29& 3.149&34 &1.991\\
\hline
& & & & & & \\
\textbf{Average}& \textbf{0.433}   & \textbf{-6}  & \textbf{0.500} & \textbf{-17}&\textbf{1.198}&\textbf{-24}&\textbf{0.723}\\
\hline
\end{tabular}
\end{center}
\label{table:USstates}
\end{table}

\begin{landscape}
\begin{table}
\label{Table:Results}
\tiny
\caption{NHS Regions: nowcasting mean absolute error per geographic location when using only death data and when using deaths and different low-latency data feeds.}
\begin{center}
\begin{tabular}{cccccccccccc}
\hline
\hline
\\[-5pt]
\multicolumn{1}{c}{\textbf{Geographic Location}} & 
\multicolumn{1}{c}{\textbf{Deaths}} & 
\multicolumn{2}{c}{\textbf{Hospital}} &
\multicolumn{2}{c}{\textbf{Twitter}} &
\multicolumn{2}{c}{\textbf{Zoe App}} &
\multicolumn{2}{c}{\textbf{111 Calls}} &
\multicolumn{2}{c}{\textbf{111 Online}}\\
\multicolumn{1}{c}{} &    
\multicolumn{1}{c}{NEES} &  
\multicolumn{1}{c}{MAE \% Diff}&
\multicolumn{1}{c}{NEES} &
\multicolumn{1}{c}{MAE \% Diff} &
\multicolumn{1}{c}{NEES} &
\multicolumn{1}{c}{MAE \% Diff} &
\multicolumn{1}{c}{NEES} &
\multicolumn{1}{c}{MAE \% Diff} &
\multicolumn{1}{c}{NEES} &
\multicolumn{1}{c}{MAE \% Diff} &
\multicolumn{1}{c}{NEES}\\
\hline
\textbf{NHS Regions}    &   &  &   \\
East of England& 0.435  & -13  & 0.419 & -7 & 0.655 & 38 & 2.908 & -15 & 0.820 &-19&0.795\\
London&   0.878   &  -36 & 0.666  & -7 & 1.163 & 131 & 3.150 & -43 & 0.750 &-47&0.754\\ 
Midlands&  0.635    & -16  & 0.466 & 13 & 0.569 & 132 & 3.330 & -19 & 0.418 &-47&0.404\\
North East and Yorkshire& 0.753   & 5  & 1.188 & -4 & 0.824 & 153 & 2.325 & -16 & 0.860 &-14&0.888\\
North West& 0.735  & -1  & 0.756 & 17 & 1.408 & 129 & 3.285 & -25 & 0.932 &-25&0.934\\
South East& 0.652   & -24  & 0.805 & -3 & 1.255 & 126 & 4.390 & 8 & 1.018 &6&0.957\\
South West& 0.545   & -69  & 0.474 & 2 & 1.432 & 160 & 2.729 & -8 & 1.617&-6&1.653\\
\hline
& & & & & & & & & & &\\
\textbf{Average}&  \textbf{0.662}    & \textbf{-22}  & \textbf{0.682} & \textbf{2} & \textbf{1.044} & \textbf{124} & \textbf{3.160} & \textbf{-17} & \textbf{0.916}&\textbf{-22}&\textbf{0.912}\\
\hline
\end{tabular}
\end{center}
\label{table:UKNHS}
\end{table}
\end{landscape}

\end{document}